\documentclass[letterpaper, final,journal]{IEEEtran}

\usepackage{cite}

\usepackage[cmex10]{amsmath} 

\usepackage{amssymb}
\usepackage{mathrsfs}

\usepackage{array}

\usepackage{enumerate}

\usepackage{xy}
\xyoption{all}

\usepackage{graphicx}

\renewcommand{\iff}{\Leftrightarrow}

\newcommand{\heen}{\fbox{\small $\Rightarrow$} }
\newcommand{\weer}{\fbox{\small $\Leftarrow$} }
\newcommand{\ox}{\otimes}

\renewcommand{\bar}{\overline}

\newcommand{\ch}{\mathrm{char}}

\DeclareMathOperator{\End}{End}

\newcommand{\<}{\langle}
\renewcommand{\>}{\rangle}
\newcommand{\x}{\times}

\newcommand{\U}{\mathcal{U}}
\newcommand{\OO}{\mathcal{O}}
\newcommand{\p}{\mathfrak{p}}

\newcommand{\R}{\mathbb{R}}
\newcommand{\C}{\mathbb{C}}

\newcommand{\Q}{\mathbb{Q}}
\newcommand{\Z}{\mathbb{Z}}

\newcommand{\cC}{\mathcal{C}}

\newcommand{\vf}{\varphi}

\newcommand{\gl}{\lambda}

\newcommand{\s}{\sigma}

\newcommand{\fp}{\mathfrak{p}}

\newcommand{\qa}[3]{({#1},{#2})_{#3}}

\newcommand{\Gal}{\mathrm{Gal}}

\newcommand{\vc}[1]{\mathbf{#1}}

\newcommand{\ma}[1]{\mathbf{#1}}

\newcommand{\bs}{\boldsymbol}

\newtheorem{lemma}{Lemma}[section]
\newtheorem{theorem}[lemma]{Theorem}
\newtheorem{propo}[lemma]{Proposition}
\newtheorem{coro}[lemma]{Corollary}

\newtheorem{defi}[lemma]{Definition}
\newtheorem{ex}[lemma]{Example}

\newtheorem{remark}[lemma]{Remark}

\newtheorem{example}[lemma]{Example}

\begin{document}


\title{Quadratic Forms and Space-Time Block Codes from Generalized Quaternion and Biquaternion Algebras}

\author{Thomas~Unger and~Nadya~Markin%
\thanks{T.~Unger  is with the School of Mathematical Sciences, University College Dublin, Belfield, Dublin~4, Ireland (e-mail: thomas.unger@ucd.ie).}
\thanks{N.~Markin is with the Division of Mathematical Sciences, School of Physical and Mathematical Sciences, Nanyang Technological University, Singapore (e-mail: nadyaomarkin@gmail.com).}}
\maketitle

\begin{abstract} In the context of space-time block codes (STBCs),  the theory of generalized quaternion and biquaternion  algebras (i.e., tensor products of two quaternion algebras) over arbitrary base fields is presented, as well as quadratic form theoretic criteria to check if such algebras are division algebras. For base fields relevant to STBCs, these criteria are exploited, via Springer's theorem, to construct several explicit infinite families of (bi-)quaternion division algebras.  These  are used to obtain new $2\x 2$ and $4\x 4$ STBCs.
\end{abstract}


\section{Introduction}

\IEEEPARstart{E}{ver} since Alamouti's work \cite{A} a decade ago, division 
algebras---both commutative (i.e., fields) and noncommutative---have played an increasingly important role in the construction of space-time block codes (STBCs) for wireless communication.

Precisely how this is done is explained very thoroughly in \cite{S1}. From a mathematical point of view, one wants to construct, for a given integer $n$, a \emph{codebook} $\cC\subset M_n(F)$ of $n\x n$-matrices over some field $F$ such that $\cC$ is \emph{fully diverse}, i.e. such that 
the difference of any two distinct elements of $\cC$ has nonzero determinant. (It is desirable that $\cC$ also satisfies additional properties, such as the \emph{full rate}, \emph{non-vanishing determinant}, \emph{good shaping} and \emph{uniform average transmitted energy} properties, cf. \cite{ORBV}, \cite{ESK}.) 

The field $F$ is usually an extension of the field $\Q$ of rational numbers, containing an appropriate signal constellation. The set $\cC\subset M_n(F)$ can be obtained as follows: choose some division algebra $D$ of dimension $n$ over $F$ (i.e., $D$ is simultaneously an $F$-vector space and a ring) and consider the left regular representation
\[\gl: D\to \End_F(D),\ a\mapsto \gl_a,\]
where $\gl_a$ is left multiplication by $a$, $\gl_a(x)=ax$ for all $x\in D$. Here $\End_F(D)$ is the algebra of $F$-linear transformations on $D$. After a choice of $F$-basis for $D$, $\End_F(D)$ can be identified with the algebra $M_n(F)$. In this way we get an embedding $\gl: D \hookrightarrow M_n(F)$.
Since all nonzero elements of $D$ are invertible, all nonzero matrices in $\gl(D)$ will have nonzero determinant. The set $\cC$ can then be chosen to be (a subset of) $\gl(D)$. The difference of any two distinct elements of $\cC$ will thus have nonzero determinant.

There are some caveats here: for coding purposes it is often more convenient to consider $D$ as a vector space over some subfield $K$, maximal with respect to inclusion ($F\subset K\subset D$). When $D$ is commutative (in other words a field), the above construction will always work with $F$ replaced by $K$. However, when $D$ is noncommutative, one should consider $D$ as a \emph{right} $K$-vector space, i.e., elements of $D$ should be multiplied on the right by scalars from $K$. It should be taken into account that in both cases $\dim_F D=(\dim_K D)^2$, resulting in $K$-matrices of smaller size.

We are only interested in noncommutative division algebras in this paper. In the STBC literature two large classes of  algebras have been studied: cyclic algebras (e.g. in \cite{S1}, \cite{ORBV}, \cite{LMG}) and crossed-product  algebras (e.g. in \cite{S2} and \cite{BO}). The latter are characterized by requiring that $K$ is a strictly maximal subfield of $D$ which is a Galois extension of $F$. Cyclic algebras are crossed-product algebras with
\emph{cyclic} Galois group $\Gal(K/F)$. We refer to \cite{L} for a  historical survey of such algebras.

Determining when precisely these algebras are \emph{division algebras} can be a nontrivial problem.
In this paper we discuss two classes of algebras,  namely (generalized) quaternion and biquaternion algebras. They have  characterizations in terms of \emph{quadratic forms}, a fact that has not fully been exploited yet
in the coding theory literature. There are particularly nice quadratic form-theoretic criteria to determine when they are division algebras.

Although many examples of codes based on quaternion algebras have appeared in the literature (first and foremost the Alamouti code), they often do so disguised as cyclic algebras (of which they are special cases), which hides their quadratic form-theoretic characterization. 
Codes based on quaternion algebras seem to have been considered explicitly for the first time in \cite{BR}.

Biquaternion algebras are tensor products of two quaternion algebras (in the older literature, the term also refers to Hamilton's quaternions considered over the field of complex numbers). They are crossed-product algebras with Galois group the Klein $4$-group (but the converse inclusion is not true),  and are thus special cases of the algebras in \cite{BO}.  Biquaternion algebras can also be characterized by a quadratic form. In particular it is relatively straightforward to determine when such an algebra is a division algebra (at least over base fields that are relevant for STBCs; it should be noted that transcendental elements are needed to achieve this, cf. Section~\ref{sec8}). It is also easy to produce biquaternion division algebras as tensor products of quaternion division algebras, addressing an issue raised in \cite[p.~3926]{S2}. Whereas quaternion algebras over a field $F$ are precisely the cyclic algebras of dimension $4$ over $F$, biquaternion algebras need not be cyclic. Indeed, the first example of a noncyclic division algebra was constructed by Albert and was in fact a biquaternion algebra, cf. \cite{L}. Explicit criteria for when a biquaternion algebra over $F$ is cyclic are presented in \cite{LLT}. If $F$ is a number field, then any biquaternion algebra is cyclic.

In this paper we use quaternion division algebras to construct $2\x 2$ STBCs and biquaternion division algebras to construct $4\x 4$ STBCs. 

This paper is organized as follows: in Section~\ref{sec.q.f} we present a brief introduction to quadratic forms. In Sections~\ref{sec3} and \ref{biquat.alg} we  respectively present the theory of quaternion and biquaternion algebras over arbitrary fields. We also give  general quadratic form theoretic criteria for such algebras to be division algebras. In Section~\ref{sec5} we obtain matrix representations of quaternion and biquaternion algebras over arbitrary fields by means of the left regular representation. 

Then we shift our attention to fields that are actually used in space-time block coding: number fields and their transcendental extensions. Such fields can be embedded into bigger fields that are complete with respect to a discrete valuation, or CDV fields for short. For CDV fields the quadratic form theoretic criterion for a (bi-)quaternion algebra to be division can be tested in a finite number of steps by means of Springer's theorem (Section~\ref{sec6}). 

In Sections~\ref{sec7} and \ref{sec8} we give examples of infinite families of non-isomorphic quaternion and biquaternion algebras over suitable fields, exploiting Springer's theorem. Together with the matrix representations from Section~\ref{sec5} these are used for the explicit construction of $2\x 2$ and $4\x 4$ STBCs. 

Whereas our $2\x 2$ codes satisfy the non-vanishing determinant (NVD) property, our $4\x 4$ codes do not. This is due to the use of transcendental elements. As a result they will generally have a poorer performance than codes that do satisfy the NVD property. On the other hand they may be useful in the construction of MMSE optimal codes, cf. \cite{RR}.

In Section~\ref{sec9} we simulate some of the $2\x 2$ codes that we obtain before coming to a conclusion in Section~\ref{sec10}.

The Appendices contain preliminaries from discrete valuation theory and number theory.

\section{Quadratic Forms}\label{sec.q.f}

For the convenience of the reader we recall some basic facts from quadratic form theory. We refer to \cite{Lam} for the details.
Let $F$ be a field and write $F^\x:=F\setminus\{0\}$. We make the technical assumption that the characteristic of $F$ is different from $2$ (in other words that $2$ is invertible in $F$). The fields considered in this paper are all extensions of the rational numbers $\Q$ and thus will satisfy this condition (but the finite field $\mathbb{F}_2$ does not, for example).

Let $V$ be a finite dimensional $F$-vector space.

\begin{defi} A \emph{quadratic form} on $V$ (over $F$) is a map $\vf: V\to F$ satisfying the following two properties:
\begin{enumerate}[(i)]
\item $\vf(\alpha \vc v)=\alpha^2 \vf(\vc v)$ for all $\vc v\in V$ and all $\alpha \in F$;
\item the map $b_\vf: V\x V\to F$ defined by
\[b_\vf(\vc v,\vc w):=\vf(\vc v+\vc w)-\vf(\vc v)-\vf(\vc w)\]
for all $\vc v,\vc w\in V$ is bilinear (and automatically symmetric).
\end{enumerate}
\end{defi}

Concretely, if we identify $V$ with $ F^n$, where $n=\dim_F V$, then $\vf$ is just a homogeneous polynomial of degree two:
for $1\leq i,j\leq n$ there exist scalars $a_{ij}\in F$ such that
\begin{equation}\label{quad.form}
\vf(\vc v)=\vf(v_1,\ldots, v_n) =\sum_{i,j=1}^n a_{ij} v_iv_j,
\end{equation}
for all $\vc v=(v_1,\ldots, v_n)\in F^n$. 

If we let $\ma A=(a_{ij})_{1\leq i,j\leq n}\in M_n(F)$ we can write \eqref{quad.form} in matrix form as
\[\vf(\vc v) = \vc v \ma A \vc v^t,\]
where $t$ denotes transposition. 

If $\psi$ is another quadratic form on $V$, with matrix $\ma B$ say, then we call $\vf$ and $\psi$ \emph{isometric} if the matrices $\ma A$ and $\ma B$ are congruent, i.e., if there exists an invertible matrix $\ma C$ such that $\ma B=\ma C\ma A \ma C^t$.  In this case we write $\vf \simeq \psi$.

One can show that a quadratic form $\vf$ is always isometric to a quadratic form $\psi$ whose matrix is diagonal,
\[\psi(\vc v)=\sum_{i=1}^n a_i v_i^2\]
for all $\vc v=(v_1,\ldots, v_n)\in F^n$ and scalars $a_1,\ldots, a_n\in F$. It is convenient to employ the short-hand notation
$\psi=\<a_1,\ldots, a_n\>$
to indicate that the matrix of $\psi$ has diagonal entries $a_1,\ldots, a_n$ and zeroes everywhere else.
Thus we can write
$\vf\simeq\<a_1,\ldots, a_n\>$.
It is customary to consider quadratic forms \emph{up to isometry}. As a consequence
the entries in $\<a_1,\ldots, a_n\>$ may be permuted and changed by nonzero squares (i.e., any $a_i$ may be replaced by $\alpha^2 a_i$ for any $\alpha \in F^\x$). 

The \emph{dimension of} $\vf$ is the integer $n$ (i.e., the dimension of the underlying vector space). If $\vf'$ is another quadratic form over $F$, say of dimension $m$, $\vf'\simeq\<b_1,\ldots, b_m\>$, then the \emph{orthogonal sum} of $\vf$ and $\vf'$,  defined by
\[\vf\perp\vf':=\<a_1,\ldots, a_n, b_1,\ldots, b_m\>\]
is a quadratic form on $F^{n+m}$ which is uniquely defined up to isometry.

The quadratic forms $\vf$ in this paper will always be \emph{nonsingular} in the sense that their \emph{determinant} $\det\vf:=a_1 a_2\cdots  a_n $ (modulo squares) is always nonzero.

 A quadratic form $\vf$ on $V$ is called \emph{isotropic over $F$} if there exists a nonzero vector $\vc v \in V$ such that $\vf(\vc v)=0$; and \emph{anisotropic over $F$} otherwise. 

Note that an anisotropic form over $F$ may become isotropic after scalar extension to a bigger field.

\begin{remark} Let $\vf=\<a_1,\ldots, a_n\>$. If there are $b_3, b_4,\ldots, b_n\in F^\x$ such that $\vf\simeq \<1,-1,b_3,b_4,\ldots, b_n\>$, then $\vf$ is clearly isotropic. The converse is also true.
\end{remark}

\section{Quaternion Algebras}\label{sec3}

\begin{defi} Let $a,b\in F^\x$. A \emph{(generalized) quaternion algebra} over $F$ is a $4$-dimensional $F$-vector space with $F$-basis $\mathscr{B}=\{1, \vc i, \vc j, \vc k\}$, whose ring structure is determined by
\begin{equation}\label{def.rel}
\vc i^2=a,\ \vc j^2=b\quad\text{and}\quad \vc{ij}=-\vc{ji}=\vc k.
\end{equation}
We denote this algebra by $\qa a b F$. Its elements are called \emph{quaternions}.
\end{defi}
It follows from the defining relations \eqref{def.rel} that $\vc k^2=-ab$ and that $\vc i$, $\vc j$ and $\vc k$ anti-commute. Addition of quaternions and multiplication of a quaternion by a scalar is component wise, while  the product of two quaternions reduces to computing sums of scalar multiples of products of the form $\bs{\xi}\bs{\eta}$ with $\bs{\xi}$, $\bs{\eta}\in\mathscr{B}$. The defining relations \eqref{def.rel} should be used to simplify these products. Note that the product of two quaternions is not commutative.

\begin{ex} Hamilton's quaternions form the algebra $\mathbb{H}=\qa{-1}{-1}{\mathbb{R}}$.
\end{ex}

Let $\vc q = q_0 + q_1\vc i+ q_2 \vc j+ q_3 \vc k \in Q:=\qa a b F$ (with $q_0, q_1, q_2, q_3\in F$). The \emph{conjugate} of $\vc q$, denoted $\bar{\vc q}$, is defined by
\[\bar{\vc q}:=q_0 - q_1\vc i- q_2 \vc j- q_3 \vc k.\]
The \emph{norm map} $N: Q \to F$ is defined by $N(\vc q):= \vc q \bar{\vc q}$ for all $\vc q \in Q$. 
A straightforward computation shows that
\begin{equation}\label{norm}
N(\vc q):= \vc q \bar{\vc q}=\bar{\vc q} \vc q =q_0^2 -a q_1^2-b q_2^2 +ab q_3^2.
\end{equation}
Thus we may regard $N$ as a quadratic form in four variables $q_0, q_1, q_2, q_3$, called the \emph{norm form} of $Q$.
Using the notation from Section~\ref{sec.q.f} we write
\[N=\<1,-a,-b,ab\>.\]

\begin{theorem} The quaternion algebra $ Q=\qa a b F$ is a division algebra if and only if its norm form $N=\<1,-a,-b,ab\>$ is anisotropic over $F$.
\end{theorem}

\begin{IEEEproof} \heen If $N(\vc q)=\vc q \bar{\vc q}$ were equal to zero for some nonzero $\vc q\in Q$, then $Q$ would contain zero divisors, contradicting the assumption that $Q$ is a division algebra.

\noindent\weer Let $\vc q$ be any nonzero element of $Q$. From $\bar{\vc q}\vc q=\vc q \bar{\vc q} =N(\vc q)\not=0$ we deduce that $\vc q$ is invertible with $\vc q^{-1}=\bar{\vc q}/{N(\vc q)}$.
\end{IEEEproof}

\begin{ex} Hamilton's quaternion algebra  $\mathbb{H}=\qa{-1}{-1}{\mathbb{R}}$ is a division algebra since its norm form $\<1,1,1,1\>$ is clearly anisotropic over $\R$. The quaternion algebra $\qa{ -1} 1 \R$ is not a division algebra since its norm form $\<1,1,-1,-1\>$ is isotropic over $\R$.
\end{ex}

\begin{remark} Let   $ Q=\qa a b F$. One can show that if $Q$ is not a division algebra, that it is then necessarily isomorphic to the full matrix algebra $M_2(F)$. This thus happens precisely when the norm form of $Q$ is isotropic. 
\end{remark}

\begin{remark} One can show that two quaternion algebras over $F$ are isomorphic as $F$-algebras if and only if their norm forms are isometric quadratic forms.
\end{remark}

\section{Biquaternion Algebras}\label{biquat.alg}

\begin{defi} A \emph{biquaternion algebra} over $F$ is a tensor product of two quaternion algebras over $F$,
\[\qa a b F \ox_F \qa c d F,\]
where $a,b,c,d\in F^\x$. Its elements are called \emph{biquaternions}.
\end{defi}

Let $\mathscr{B}_1:=\{1, \vc i_1, \vc j_1, \vc k_1\}$ and $\mathscr{B}_2:=\{1, \vc i_2, \vc j_2, \vc k_2\}$ be $F$-bases for $\qa a b F$ and $\qa c d F$, respectively. The elements of $\mathscr{B}_1$ are assumed to commute with the elements of $\mathscr{B}_2$.
The ordered set
\[\mathscr{B}:=\{\boldsymbol{\xi} \ox \boldsymbol{ \eta} \mid \boldsymbol{ \xi} \in \mathscr{B}_1,\ \boldsymbol{ \eta} \in \mathscr{B}_2\}\] 
is an $F$-basis for $B:=\qa a b F \ox_F \qa c d F$. Thus $\dim_F B=16$ and we can express a biquaternion $\vc x\in B$ uniquely in the form
\[\vc x =\sum_{\bs{\xi}\in \mathscr{B}_1} \sum_{\bs{\eta}\in \mathscr{B}_2} x_{\bs{\xi},\bs{\eta}}\, \bs{\xi}\ox\bs{\eta},\]
with $x_{\bs{\xi},\bs{\eta}} \in F$. Addition of biquaternions and multiplication of a biquaternion by a scalar is component wise, while  the product of two biquaternions reduces to computing sums of scalar multiples of products of the form
\[(\bs{\xi}\ox \bs{\eta})(\bs{\xi}' \ox \bs{\eta}') = \bs{\xi\xi}' \ox \bs{\eta \eta}',\]
where $\bs{\xi}, \bs{\xi}'\in \mathscr{B}_1$ and $\bs{\eta}, \bs{\eta}'\in \mathscr{B}_2$. The defining relations \eqref{def.rel} should be used to simplify these products. For example, 
\begin{align*}
(\vc i_1\ox \vc j_2) (\vc i_1 \ox \vc k_2)&= \vc i_1^2 \ox \vc j_2 \vc k_2= a \ox \vc j_2 \vc i_2 \vc j_2=
a\ox (-\vc i_2 \vc j_2^2)\\
&=a\ox (-d \vc i_2) =-ad(1\ox \vc i_2).
\end{align*}
Note that the product of two biquaternions is not commutative.

\begin{defi} Let $B:=\qa a b F \ox_F \qa c d F$. The six-dimensional quadratic form
\[\vf=\<a,b,-ab -c,-d, cd\>\]
over $F$ is called the \emph{Albert form} of $B$.
\end{defi}

This form is named after Adrian A.~Albert who carried out much of the pioneering work on the structure of algebras in the 1930s.

\begin{theorem} The biquaternion algebra $ B=\qa a b F \ox_F \qa c d F$ is a division algebra if and only if its Albert form $\vf=\<a,b,-ab, -c,-d, cd\>$ is anisotropic over $F$.
\end{theorem}

For a proof  of this theorem we refer to \cite[III, Theorem~4.8]{Lam}. 

\begin{remark} If the Albert form $\vf$ of $B$ is isotropic over $F$, then one can show that
$ B$  isomorphic to the full matrix algebra $M_4(F)$ if and only if $\vf$ is isometric to $\<1,-1,1,-1,1,-1\>$ and also that $B\cong M_2(D)$ for some quaternion division algebra $D$ if and only if $\vf\simeq \<1,-1,e,f,g,h\>$  with $\<e,f,g,h\>$ anisotropic. 
\end{remark}

\begin{remark} One can show that two biquaternion algebras over $F$ are isomorphic as $F$-algebras if and only if their Albert forms are \emph{similar}, i.e., isometric up to a multiplicative constant from $F$.\end{remark}

\section{Matrix Representations}\label{sec5}

\subsection{The Quaternion Case}

Let $Q=\qa a b F$ be a quaternion division algebra with $F$-basis $\{1, \vc i, \vc j, \vc k\}$, so that $\vc i^2=a$ and $\vc j^2=b$. The field $K:=F(\sqrt{a})=F(\vc i)$ is a maximal subfield of $Q$ with Galois group $\Gal(K/F)=\{1, \s\}$ where $\s$  (the nontrivial automorphism of $K$ that fixes $F$) is determined by $\s(\sqrt{a})=-\sqrt{a}$. 

Let $\vc x = q_0 + q_1\vc i+ q_2 \vc j+ q_3 \vc k \in Q$ (with $q_0, q_1, q_2, q_3\in F$). Since
\[\vc x = q_0 + q_1\vc i+ q_2 \vc j+ q_3 \vc k = (q_0 +q_1 \vc i) + \vc j (q_2-q_3 \vc i)=x_0 +\vc j x_1\]
with $x_0=q_0+q_1\vc i$ and $x_1=q_2 - q_3 \vc i$ in $K$, we can write
\[Q=F(\vc i)\oplus \vc j F(\vc i)=K\oplus \vc j K.\]
Hence $Q$ can be considered as a two-dimensional right $K$-vector space with $K$-basis $\{1,\vc j\}$.
We thus multiply elements $\vc x$ from $Q$ with scalars $\alpha$ from $K$ on the \emph{right} to get $\vc x \alpha$. In computations we may bump into expressions of the form $\alpha \vc x$ with scalars on the left. They can be brought into the correct form by means of the ``hopping over'' rule
\begin{equation}\label{hop1}
\alpha \vc x = \vc x \s(\alpha)
\end{equation}
(which can be verified to be true  by direct computation).

\begin{theorem}\label{quatmat} \[Q\cong \Biggl\{ 
\begin{bmatrix}
x_0 & b\s(x_1) \\
x_1 & \s(x_0)
\end{bmatrix} \,\Bigg\vert\, x_0, x_1 \in K \Biggr\}\]
\end{theorem}

\begin{IEEEproof} Let $\lambda :Q\hookrightarrow \End_K(Q),\ \vc x\mapsto \lambda_{\vc x}$ be the left regular representation. We can identify $\End_K(Q)$ with $M_2(K)$ by mapping $\lambda_{\vc x}$ to its matrix with respect to the $K$-basis $\{1,\vc j\}$ of $Q$, which we will denote by $\ma X$. Let $\vc x= x_0+\vc j x_1\in Q$. Since
\[
\begin{aligned}
\lambda_{\vc x} (1) &= \vc x = x_0 + \vc j x_1,\\
\lambda_{\vc x} (\vc j) &= \vc x \vc j = (x_0+\vc j x_1)\vc j = x_0\vc j+ \vc j x_1 \vc j = \vc j \s(x_0) +\vc j^2 \s(x_1)\\
&=b\s(x_1) + \vc j\s(x_0)
\end{aligned}
\]
(where we used the ``hopping over'' rule \eqref{hop1} and the fact that $\vc j^2=b$),
we obtain
\[\ma X= \begin{bmatrix}
x_0 & b\s(x_1) \\
x_1 & \s(x_0)
\end{bmatrix}.\]
Since the map $\vc x \mapsto \ma X$ yields an isomorphism of $Q$ with its image inside $M_2(K)$ we are done.
\end{IEEEproof}

\begin{ex} Letting $F=\R$, $K=\C$ and $Q=\qa{-1}{-1}{\R}$ we obtain the Alamouti code.
\end{ex}

More examples will be constructed in Section~\ref{sec7}.

\begin{remark} Note that
\[\det\ma X = x_0\s(x_0) - bx_1 \s(x_1)= q_0^2 -aq_1^2-bq_2^2+abq_3^2=N(\vc x),\]
confirming once again that invertible elements of $Q$ (i.e., elements with nonzero norm) correspond to invertible matrices.
\end{remark}

\begin{remark} The field $K=F(\sqrt{a})$ is a so-called \emph{splitting field} for $Q$ in the sense that $Q\ox_F K \cong M_2(K)$. 
\end{remark}

\subsection{The Biquaternion Case}
Let $B:=\qa a b F \ox_F \qa c d F$ be a biquaternion division algebra. 
Let  $\mathscr{B}_1$, $\mathscr{B}_2$ and $\mathscr{B}$ be as in Section~\ref{biquat.alg}. Then $\vc i_1^2=a$, $\vc j_1^2=b$, $\vc i_2^2=c$ and $\vc j_2^2=d$. The field $K:=F(\sqrt{a}, \sqrt{c})$ is a maximal subfield of $B$ with Galois group $\Gal(K/F)=\{1, \s, \tau, \s\tau\}$ where $\s$  and $\tau$ are determined by $\s(\sqrt{a})=-\sqrt{a}$ and $\tau(\sqrt{c})=-\sqrt{c}$, respectively, and where $\s\tau$ denotes the composition of $\s$ and $\tau$.

The field $K$ is a biquadratic extension of the field $F$. The following diagram illustrates the relationship between $K$, $F(\sqrt{a})$, $F(\sqrt{c})$ and $F$:
\[\begin{aligned}
\xymatrix{
& K=F(\sqrt{a}, \sqrt{c}) & \\
F(\sqrt{a})\ar@{-}[ur]^-{2}  \ar@{-}[dr]_-{\sigma} & & F(\sqrt{c}) \ar@{-}[ul]_-{2} \ar@{-}[dl]^-{\tau}\\
& F &
}
\end{aligned}\]
A $K$-basis for $B$ is given by $\{1\ox 1, 1\ox \vc j_2, \vc j_1\ox 1, \vc j_1\ox \vc j_2\}$.
One can easily verify that $B$ is a right $K$-vector space which can be identified with 
\[K\oplus \vc j_1 K \oplus \vc j_2 K \oplus \vc j_1 \vc j_2 K.\]
This time there are several ``hopping over'' rules for multiplication by scalars on the left, determined by
\begin{equation}\label{hop2}
\alpha \vc j_1 = \vc j_1 \s(\alpha),\  \alpha \vc j_2 = \vc j_2 \tau(\alpha) \text{ and } \alpha \vc j_1 \vc j_2= \vc j_1 \vc j_2\s\tau(\alpha)
\end{equation}
for all $\alpha\in K$.

\begin{theorem} \label{biquatmat}
\[B\cong \left\{ 
\begin{bmatrix}
x_0 & b\s(x_1) & d\tau(x_2) & bd \s\tau (x_{12})\\
x_1 & \s(x_0) & d\tau(x_{12}) & d \s\tau (x_{2})\\
x_2 & b\s(x_{12}) & \tau(x_{0}) & b \s\tau (x_{1})\\
x_{12} & \s(x_2) & \tau(x_{1}) &  \s\tau (x_{0})
\end{bmatrix}\right\},\]
where $x_0, x_1, x_2, x_{12} \in K$.
\end{theorem}

The proof is similar to the proof of Theorem~\ref{quatmat} but this time the ``hopping over'' rules \eqref{hop2} should be used.

\section{Springer's Theorem}\label{sec6}

So far we have described quaternion and biquaternion algebras over arbitrary fields (in which $2$ is invertible) and shown how to compute their matrix representations. In addition we have seen that a given quaternion or biquaternion algebra is a \emph{division} algebra if and only if its norm form or Albert form, respectively, is anisotropic. 

The next question is how to determine when a quadratic form over a given field is actually anisotropic. In general this is a difficult problem, but for the fields in which we are interested we are in luck! The idea is to embed them into bigger fields that are complete with respect to a discrete valuation. For such fields there exists an easy criterion for anisotropy: Springer's theorem. If the quadratic form is anisotropic over the bigger field, then it was already anisotropic over the smaller field. Applying this to a norm form or Albert form  we get that if a quaternion or biquaternion algebra is a division algebra over the bigger field, then it was already division over the smaller field.

We will now present the details of this approach. Our fields of interest are:
\begin{itemize}
\item number fields $F$ (i.e., finite extensions of the field $\Q$ of rational numbers);
\item rational function field extensions of $F$ of the form $F(x,y)$.
\end{itemize}
These fields are equipped with a \emph{discrete valuation}. 
We embed them into bigger fields:
\begin{itemize}
\item $F\hookrightarrow F_\fp$: $\fp$-adic field;
\item $F(x,y)\hookrightarrow F(\!(x)\!)(\!(y)\!)$: iterated Laurent series field.
\end{itemize}
These bigger fields are \emph{complete with respect to a discrete valuation} (or CDV fields for short). They are the completions of $F$ and $F(x,y)$, respectively, for an appropriate discrete valuation. We refer to  Appendix~\ref{appA}  for more background.

\begin{theorem}[Springer's theorem on CDV fields]\label{Springer}
Let $F$ be a CDV field with residue field $\bar{F}$, uniformizer $\pi$ and group of units $\U$. Assume that $F$ is non-dyadic (i.e., $\ch(\bar F)\not=2$). Let $\vf$ be a quadratic form of dimension $n$ over $F$. 
Since the entries of $\vf$ are defined up to squares we may
write
\[\vf=\vf_1 \perp \<\pi\> \vf_2\]
where $\vf_1=\<u_1,\ldots, u_r\>$, $\vf_2=\<u_{r+1},\ldots, u_n\>$ and $u_i\in \U$ for $1\leq i \leq n$. Then $\vf$ is anisotropic over $F$ if and only if the ``first residue form'' $\bar\vf_1=\<\bar u_1,\ldots, \bar u_r\>$ and the ``second residue form'' $\bar\vf_2=\<\bar u_{r+1},\ldots, \bar u_n\>$ are anisotropic over $\bar F$.
\end{theorem}

We refer to \cite[Chapter VI, \S1]{Lam} for more information on Springer's theorem.

\section{$2\times 2$ STBCs from Quaternion Algebras}\label{sec7}

Using Springer's theorem we construct infinite families of quaternion division algebras over number fields. They give rise to $2\times 2$ STBCs by means of the matrix representation in Theorem~\ref{quatmat}. We refer to  Appendix~\ref{appB}  for more background on number fields.

\begin{theorem}
Let $F$ be a number field, $\frak p$ a prime ideal of $\OO_F$ with corresponding $\p$-adic valuation $v_\p$ and $\pi$ an element of $F$ such that $v_\p(\pi) = 1$. Then for any element $b$ in the ring of integers $\mathcal O_F$ of $F$, which is not a square modulo $\frak p$,  the quaternion algebra $\qa \pi b F$ is a division algebra. 
\label{mainprop}
\end{theorem}

\begin{IEEEproof} 
We show that the norm form $N = \<1,-\pi,-b,\pi b\>$ is anisotropic over $F$.
Consider the $\p$-adic valuation $v_\p$ on $F$. 
We obtain the following decomposition of $N$ in the completion $F_\p$: $$N = \varphi_1 \perp \<\pi\> \varphi_2, $$ where $\varphi_1 = \<1, -b\>$ and $ \varphi_2 = \<-1,b\>$.  Let $\bar \varphi_i$ denote the reduction of $\varphi_i$ modulo $\p$. Since we chose $b$ to be a non-square in $\OO_F/\p$, the equation $x^2b = y^2$ has no nontrivial solutions in $\OO_F/\p$, therefore both
 $\bar \varphi_1$ and $\bar \varphi_2$ are anisotropic in the residue field $\bar{F_\p}$. By Springer's theorem it follows that $N$ is anisotropic over $F_\p$. Since $F$ is embedded in $F_\p$, any nontrivial zero of $N$ over $F$ would give rise to a nontrivial zero over $F_\p$. It follows that $N$ is anisotropic over $F$. Hence $\qa \pi b F$ is a division algebra.
\end{IEEEproof}

\begin{coro}
Let $F$ be a number field. There are infinitely many quaternion division algebras $\qa \pi b F$.
\end{coro}

\begin{IEEEproof}
For any number field $F$, there are infinitely many primes $\p$ in $\OO_F$. Since any element of $\OO_F$ is contained in only finitely many primes, it can only be a uniformizer for finitely many prime ideals. Thus we obtain
infinitely many elements $\pi = \pi_\p$. Having picked $\pi$, we can always find $b \in \OO_F$ which
 is non-square modulo $\pi$. This is because the map $x\mapsto x^2$ in $\OO_F/\p^\times$ is never surjective (since $\p$ is not a prime above $2$, it has a nontrivial kernel containing $-1$).
\end{IEEEproof}

From Theorem~\ref{quatmat} it follows that for
$\pi$ and $b$ as in Theorem~\ref{mainprop} we obtain (after transposition) the infinite codebook
\begin{equation}\label{codebook}
\cC(Q):=\Biggl\{\begin{bmatrix}
\alpha+\beta\sqrt{\pi} & \gamma+\delta\sqrt{\pi} \\
b(\gamma-\delta\sqrt{\pi}) & \alpha-\beta\sqrt{\pi}
\end{bmatrix} \,\Bigg\vert\, \alpha, \beta, \gamma, \delta \in \OO_F \Biggr\}
\end{equation}
from the algebra $Q=\qa \pi b F$.
The code $\cC(Q)$ has the following properties: it is full rate (since 4 
transmitted signals
are used to transmit 4 information symbols); it is full rank (since it comes from a division algebra); as long as $|b|^2=1$ it also has uniform average energy per antenna. For a codebook $\cC$ constructed from a division algebra, let 
\[\delta_{\min}(\cC):=\inf_{\ma{X}\not=\ma{X}'\in \cC}|\det(\ma{X}-\ma{X}')|^2= 
\inf_{0\not=\ma{X}\in \cC}|\det(\ma{X})|^2\] 
denote its minimum determinant.

\begin{theorem} \label{thm7.3}
Let $F=\Q$ or $F=\Q(\sqrt{-d})$ for some square-free positive integer $d$, let $Q=\qa \pi b F$ be the quaternion division algebra from Theorem~\ref{mainprop}, and let $K/F$ be a maximal commutative subfield of $Q$ with ring of integers $\OO_K$. Then the quaternionic code $\cC(Q)$, constructed from $Q$ and with entries from $\OO_K$ satisfies the non-vanishing determinant property.
\end{theorem}

\begin{IEEEproof} This follows immediately from \cite[Example~17.9]{Be-Og2}, where it is shown that for a quaternion algebra $Q=(a,b)_F$ we have
$\delta_{\min}(\cC(Q))\geq 1/|b_d|^2$, where $b=b_n/b_d$ with $b_n, b_d\in \OO_F$, $b_d\not=0$.  Since we choose $b \in \OO_F$ in Theorem~\ref{mainprop}, we obtain $\delta_{\min}(\cC(Q))\geq 1$.
\end{IEEEproof}

\begin{remark}
As a consequence these codes achieve the diversity-multiplexing gain tradeoff of Zheng and Tse \cite{ZT}, as explained in 
\cite{EKPKH}.
\end{remark}

In the examples below we illustrate how to use Theorem~\ref{mainprop}.
We start with the simplest case when the number field $F$ is $\Q$. 

\begin{example} The quaternion algebra $\qa 3 {-1} \Q$ is a division algebra.
\end{example}

\begin{IEEEproof}
The ideal $\p$ in this case is simply $3\Z$, with $\pi=3$. Let $\bar x$ denote the reduction of $x$ modulo $\p$. The squares in $\Z/3\Z$ are $\{\bar{0}, \bar{1}\}$. Since $-1 \equiv 2 \mod 3$, it is not a square, hence the conditions of Theorem~\ref{mainprop} are satisfied. We conclude that $\qa 3 {-1} \Q$ is a division algebra. 
\end{IEEEproof}

Note that instead of $-1$ we could use any element congruent to $2$ modulo $3$. Also note that $\qa 3 {-1} \Q$ does not remain a division algebra upon extending scalars to $\Q(i)$ since $-1=i^2$ is a square in $\Q(i)$.

Now let $F=\Q(i)$.

 \begin{propo}\label{prop7.6}
Let $p$ be an odd prime such that $p = x^2 + y^2$ is a sum of two squares and $p \not \equiv 1 \bmod 8$. Then $\qa {x + iy} i {\Q(i)}$ and $\qa {x - iy} i {\Q(i)}$ are both division algebras. 
\end{propo}
 \begin{IEEEproof}
The conditions on the prime $p$ imply that the ideal $(p)$ splits into two prime ideals $(x+iy)(x-iy)$ in $\Z[i]$. Since the inertial degree of each of the prime ideals dividing $p\Z[i]$ is 1, we have $\Z[i]/(x\pm iy) \cong \Z/p\Z$. Since $p \not \equiv 1 \bmod 8$, there is no element of order 8 in $\Z/p\Z^\times$, and hence there is no element of order $8$ in each of $\Z[i]/(x \pm iy)$. This implies that $i$ is not a square in $\Z[x]/(x \pm iy)$. By Theorem~\ref{mainprop} the algebra  $\qa {\pi} i {\Q(i)}$ is a division algebra, for each $\pi =x \pm iy$.
 \end{IEEEproof}

\begin{example}\label{sumsquares}
It can be shown that any prime $p \equiv 1 \bmod 4$ is a sum of two squares. 
Consider $p=5 = 1^2+2^2$, giving the decomposition  $5 = (1+2i)(1-2i)$ in $\Z[i]$. 
By Proposition~\ref{prop7.6} the algebra $\qa \pi i {\Q(i)}$ is a division algebra
for each $\pi = {1\pm 2i}$.  Since also $5=(2+i)(2-i)$ in $\Z[i]$, we obtain the quaternion division algebras $\qa {2\pm i} i {\Q(i)}$.
Some more values of $\pi$ that give rise to a division algebra 
are:  $\pi = 2\pm 3i$ corresponding to the prime $13 = 2^2 + 3^2$, $\pi = 1 \pm 4 i$ corresponding to the prime $17 = 1^2 + 4^2$, etc. 
\end{example}

In the next example we choose a prime ideal $\p$ of $\Z[i]$ with  nontrivial inertial degree. 

\begin{example} Let $\p = 7 \Z[i]$. There are $24$ nonsquare classes in $\Z[i]/\p$ and for any $b \in \Z[i]$ which is not a square modulo $\p$, we have $\qa 7 b {\Q(i)}$ is a division algebra.
\end{example}

\begin{IEEEproof}
By Theorem~\ref{pInZi}, the ideal $\p=7 \Z[i]$ is the principal ideal generated by $7$ and its inertial degree over $7\Z$ is $2$.
Hence  
the number of elements in $\Z[i]/\p$ is $7^2=49$. 
As an additive group the residue field $\Z[i]/\p$ is isomorphic to $\Z/7\Z \oplus \Z/7\Z$: it has $49$ elements of the form $\bar a + \bar b i$ with  $\bar a, \bar b \in \Z/7\Z$. Consider the map $s: \Z[i]/\p^{\times} \rightarrow \Z[i]/\p^{\times}$ of the multiplicative group  $\Z[i]/\p^{\times}$,  which sends every element to its square. The kernel of $s$ is of order $2$ and therefore the order of the  image is $|\Z[i]/\p^{\times}|/2 =24$. 
More precisely, the image consists of $24$ nonzero squares in $\Z[i]/\p$, which all have the form $(\bar x + \bar y i)^2 = (\bar x^2 - \bar y ^2)+ i 2 \bar{xy}$, where $x,y \in \Z$ and the bar denotes  reduction modulo $7$. This leaves $24$ choices for the congruence class of $b$ which satisfy the condition of Theorem~\ref{mainprop}. 
 \end{IEEEproof}

Consider a quaternion division algebra $Q=(\pi,b)_F$ as in Theorem~\ref{mainprop}. Since the quaternion algebra $Q'=(b,\pi)_F$ is isomorphic to $Q$, it is also a division algebra and so yields a quaternionic code $\cC(Q')$. (A priori the codes $\cC(Q)$ and $\cC(Q')$ need not have the same performance.) 

\begin{example}\label{ex7.8} 
The quaternionic code from Belfiore--Rekaya \cite{BR}.
Consider the quaternion algebra $Q=\qa {1+2i} i {\Q(i)}$, which is a division algebra by 
Example~\ref{sumsquares}. Consider the isomorphic division algebra  $Q'=\qa i {1+2i}  {\Q(i)}$. The quaternionic code $\cC(Q')$ was introduced in \cite{BR}. Since 
$|1+2i|^2=5\not=1$ the energy at the transmit antennas is unbalanced. This was remedied in \cite{BR} by changing the codewords from 
\begin{equation}
\begin{bmatrix}
\alpha+\beta\sqrt{i} & \gamma+\delta\sqrt{i} \\
(1+2i)(\gamma-\delta\sqrt{i}) & \alpha-\beta\sqrt{i}
\end{bmatrix}
\end{equation}
to
\begin{equation}\label{BRcode}
\begin{bmatrix}
\alpha+\beta\sqrt{i} & \sqrt{1+2i}(\gamma+\delta\sqrt{i}) \\
\sqrt{1+2i}(\gamma-\delta\sqrt{i}) & \alpha-\beta\sqrt{i}
\end{bmatrix},
\end{equation}
where $\alpha, \beta, \gamma, \delta \in \Z[i]$.
This does not change the codeword determinants. We will denote the code with codewords as in \eqref{BRcode} by $\cC_{BR}$. It was shown in \cite{BR} that this code performs better than the $\boldsymbol{B}_2$-code of \cite{DTB}, a version of the $2\x 2$ TAST code of \cite{EGD}.
\end{example}
\begin{remark} \label{rem7.9}
The cyclic algebra on which the Golden Code $\cC_G$ \cite{B-R-V} is built is isomorphic to the quaternion division algebra $\qa 5 i {\Q(i)}$, which is not in the class of quaternion algebras described in Theorem~\ref{mainprop} since $i$ is a unit in $\Z[i]$ and $5$ is not prime in $\Z[i]$.
\end{remark}

\section{$4\times 4$ STBCs from Biquaternion Algebras}\label{sec8}

The following theorem can be used to construct $4\times 4$ STBCs by means of the matrix representation in Theorem~\ref{quatmat}.

\begin{theorem}\label{thm8.1}
Let $K = F(x,y)$, where $F$ is a number field, and $x,y$ are transcendental over $F$.   Let $B = \qa a x K \ox \qa b y K$. 
If $a,b \in F^\x$ are such that $a, b, ab \not \in F^{\times 2}$, then $B$ is a division algebra over $K$. 
\end{theorem}

\begin{IEEEproof} The biquaternion algebra $B$ will be a division algebra over $K$ if and only if its Albert form 
\[\vf = \< a,x,-ax,-b,-y, by\>\]
is anisotropic over $K$. We will show that this is the case by an iterated application of Springer's Theorem. 

Consider the embedding of $K=F(x,y)$ into the iterated Laurent series field $F(\!(x)\!)(\!(y)\!)$.  (Note that $F(\!(x)\!)(\!(y)\!)\not=F(\!(y)\!)(\!(x)\!)$.). Let $v_y$  denote the $y$-adic valuation of $F(\!(x)\!)(\!(y)\!)$. We obtain the decomposition 
\[\vf = \vf_1 \perp \<y\>\vf_2,\]
where $\vf_1 = \<a,x,-ax,-b\>$ and $\vf_2 = \<-1,b\>$.  The form $\vf$ will be anisotropic over 
$F(\!(x)\!)(\!(y)\!)$ if and only if the forms $\bar\vf_1$ and $\bar\vf_2$ are anisotropic over  $F(\!(x)\!)$, the residue field of $v_y$. The form $\bar\vf_2$ is clearly anisotropic over $F(\!(x)\!)$ by our assumption. We can decompose the form $\bar\vf_1$ as
\[\bar\vf_1 = \psi_1 \perp \<x\> \psi_2,\]
where
$\psi_1 = \<a, -b\>$ and $\psi_2 = \<1,-a\>.$ By our assumption the residue forms $\bar\psi_1$ and $\bar\psi_2$ are clearly anisotropic over $F$ so that $\bar\vf_1$ is anisotropic over $F(\!(x)\!)$.
In conclusion, the form $\vf$ is anisotropic over $F(\!(x)\!)(\!(y)\!)$, and thus certainly anisotropic over the smaller field $K$.
\end{IEEEproof}

\begin{remark} In order to have uniform average energy at the transmitters one chooses
transcendental elements $x,y$ on the unit circle.  
\end{remark}
 
\begin{remark} The use of transcendental elements cannot be avoided here since biquaternion algebras are never division algebras when their base field is a number field, cf. \cite[p.~342]{Lam}. As a consequence codes based on these algebras do not satisfy the non-vanishing determinant property.
\end{remark}

\begin{example}  Note that the proof of Theorem~\ref{thm8.1} holds \emph{a fortiori} when $a,b,x,y$ are all chosen to be independent transcendental numbers over $F$.
In fact, an example of such a code is the code in \cite[Example~3.3]{RR}, constructed from a crossed product algebra. It is a biquaternion code where $F=\Q(i)$ and $a,b,x,y$ are all chosen to be transcendental numbers on the unit circle.  This code was shown to be MMSE optimal in \cite{RR}.
\end{example}

\begin{remark} The $4\times 4$ codes in \cite{BO} obtained from crossed product algebras
are examples of codes outside the class of biquaternion codes. 
\end{remark}

\section{Simulations}\label{sec9}

\begin{figure}[!t] 
\centering 
\includegraphics[width=.49\textwidth]{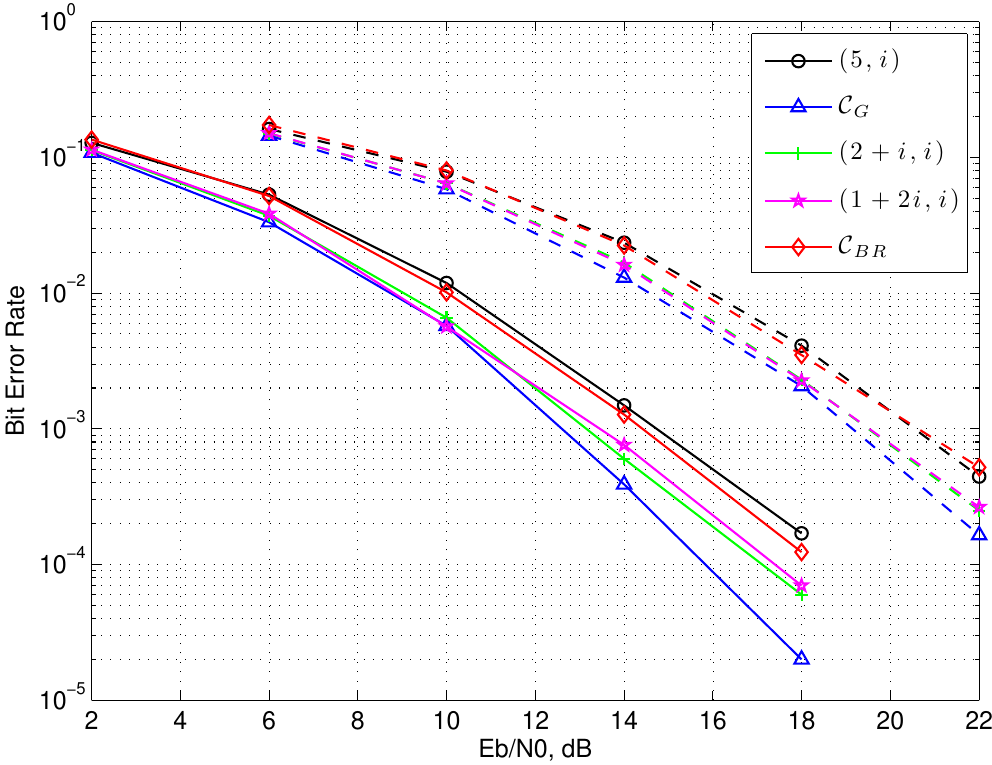} 
\caption{Simulation results for $2\x2$ codes: 4-QAM (solid lines), 16-QAM (broken lines)} 
\label{fig_sim} 
\end{figure}

Fig.~\ref{fig_sim}  shows the performance of  $2\times 2$ quaternionic codes for the Rayleigh fading channel for 4-QAM (solid line) and 16-QAM  (broken line) constellations. Decoding was done with the Sphere Decoder. The codes are  based on the following quaternion algebras:   $\qa {2+i} i {\Q(i)}$ from Example~\ref{sumsquares}, $\qa {1+2i} i {\Q(i)}$,  from Example~\ref{ex7.8} and
$\qa {5} i {\Q(i)}$ from Remark~\ref{rem7.9}. The Golden Code $\cC_G$ from \cite{B-R-V} 
and the code $\cC_{BR}$ from \cite{BR} are also present for comparison purposes.
All codewords in a quaternionic code coming from $Q=(a,b)_F$ have been scaled by a factor $1/\sqrt{P}$ where $P= (1+|\sqrt{a}|^2)(3+|b|^2)/4$ to ensure that the average power transmitted by each antenna per channel use is one, cf. \cite[p.~3918]{S2}. Observe that the codes based on the quaternion algebras  $\qa {2+i} i {\Q(i)}$ and $\qa {1+2i} i {\Q(i)}$ perform better than the code $\cC_{BR}$, but not as well as the Golden Code $\cC_G$. One can construct more codes that behave in this way (e.g. the code based on $\qa {-1+2i} i {\Q(i)}$ has the same performance as the code based on $\qa {1+2i} i {\Q(i)}$).  Also observe the enormous performance difference between the Golden Code $\cC_G$ and the quaternion code based on $\qa {5} i {\Q(i)}$, which underlies the Golden Code. This is due to the fact that the Golden Code has good shaping and the other one does not. Since the quaternionic codes introduced in this paper all outperform the code based on $\qa {5} i {\Q(i)}$ one can therefore ask the question if they can be modified so as to be comparable to the Golden Code.

\section{Conclusion}\label{sec10}
We discussed the theory of (generalized) quaternion and biquaternion algebras in the context of $2\times 2$ and $4\x 4$ STBCs, respectively, and showed how the use of associated quadratic forms (the norm form and Albert form, respectively) allows us to construct division (bi)-quaternion algebras in a uniform manner, and from them STBCs. In this we way we recover some known codes from the literature (the $2\x 2$  code from \cite{BR}, the $4\x 4$  code from \cite{RR}) and obtain some new $2\x 2$ codes which perform better than the $2\x 2$ TAST code from \cite{EGD} in simulations, but not as well as the Golden Code from \cite{B-R-V}. Future work will include optimizing the codes presented here.

\appendices

\section{Preliminaries from Discrete Valuation Theory}\label{appA} 
We refer to \cite[Chapter 1]{FV} for the details. 
 Let $F$ be a field. A \emph{ discrete valuation} on $F$ is a map $$v: F \rightarrow \Z \cup \{\infty\}$$ with the following properties:
\begin{itemize}
\item 
$v(a) = \infty \iff a = 0$,

\item $v(ab) = v(a)+v(b)$,
 
\item $v(a+b) \geq \min(v(a), v(b))$.

\end{itemize}

Any discrete valuation $v$ on $F$ induces an \emph{absolute value} $|a| := c^{v(a)}$, where $0 < c < 1$ is a real constant. Two absolute values $|\cdot|_1, |\cdot|_2 $ are said to be equivalent if there is a constant $t$ such that $|a|_1= |a|_2^t$ for all $a \in F$. Let $|\cdot|_v$ denote the equivalence class of absolute values induced by $v$.  If $v$ is a discrete valuation on $F$, then $F$ has a completion  $F_v$ with respect to  $|\cdot|_v$. This completion  is unique up to an isomorphism over $F$. 
Let $v$ also denote the extension of the valuation $v$ to $F_v$.

Given a valuation $v$ on $F$, let 
\begin{align*}
\OO_v &:= \{ a \in F_v \mid v(a) \geq 0\},\\
\mathfrak{m}_v &:= \{a \in F_v \mid v(a) > 0 \}, \text{ and}\\ 
 \U_v &:= \{a \in F_v \mid v(a) = 0 \}. 
\end{align*} 
 The objects just defined are the \emph{valuation ring}, the \emph{maximal ideal} and the \emph{group of units} of $F_v$, respectively. 
The ring $\OO_v$ is a local ring and $\frak m_v$ is its unique maximal ideal. The \emph{residue field}  $\bar F_v$ of $v$ is defined to be $\OO_v/{\frak m_v}$. 

An element $\pi$ of $F_v$ such that $v(\pi)=1$ is called {\emph{a uniformizer}}. Any two uniformizers differ by multiplication by a unit. Any element $x \in F_v^\times$ has the form $x = \pi^i u$, where $i$ is an integer and $u \in  \U_v$ is a unit. 
This gives us the following isomorphism of the multiplicative group of $F_v$: $$F_v^\times \cong  {\Z} \times \U_v.$$

Below we give some examples of discrete valuations. 

\begin{example}
Consider a prime ideal $p\Z$ of $\Z$. Any nonzero rational number can be expressed as $ {p^a x}/{y}$, where $a$, $x$ and $y$ are integers and $p \nmid xy$. Define the \emph{$p$-adic valuation} on $\Q$ by $v_p : {p^a x}/{y} \mapsto a$.  Then 
$\Q_p$ denotes the completion of $\Q$ with respect to the $p$-adic valuation and $\Z_p$ denotes the valuation ring of $\Q_p$, also known as the ring of $p$-adic integers.  The residue field $\bar \Q_p \cong \Z_p/p\Z_p \cong \mathbb F_p$ is a finite field of order $p$. 
\end{example}

Similarly we can define \emph{$\p$-adic valuations} on  number fields, as in the next example. 

\begin{example}
Let $F$ be a number field with  ring of integers $\OO_F$. Any prime ideal $\p$ of $\OO_F$ induces a discrete valuation on $F$. Namely, let $v_\p$ map a nonzero element $x$ of $\OO_F$ to the largest integer $i$, such that $x \in \p^i $, and let $v_\p(0) = \infty$. Extend this map to $F$ by letting $v_\p(x/y) := v_\p(x) - v_\p(y)$. The map just defined is a discrete valuation on $F$. We let $F_\p$  denote $F_{v_\p}$, the completion of $F$ with respect to $v_\p$. Similarly let $\OO_\p$ and $\U_\p$ denote the ring of integers of $F_\p$ (also called $\p$-adic integers) and the group of units of $\OO_\p$ (called $\p$-adic units), respectively. 
\end{example}

\begin{example} \label {ils}
Let $F$ be a field and let $K = F(x)$ . For 
\[K\ni f(x)= \sum_{i = m}^{N} a_ix^i, \qquad a_m \not = 0,\ m\in\Z\] 
define $v_x(f) = m$. 
The completion of $K$ with respect to $v_x$ is the field of  Laurent series, denoted by $F(\!(x)\!)$. Its elements are of the form $ \sum_{i = m}^{\infty} a_ix^i$ with $m\in\Z$. The valuation ring of $F(\!(x)\!)$ with respect to $v_x$ is the ring of formal power series $F[\![x]\!]$, whose elements are  of the form $ \sum_{i = 0}^{\infty} a_ix^i$. Its residue field is $F$.
\end{example}

\section{Preliminaries from Number Theory}\label{appB}

Next we look at how prime ideals behave in extensions of number fields. Let $F$ be a number field, i.e., a finite extension of $\Q$. Let $K$ be a finite extension of $F$. If $\p$ is a prime
ideal of $\OO_F$, then $\p \OO_K$ is an ideal of $\OO_K$ with unique
factorization into prime ideals of $\OO_K$:
\[\p \OO_K = \frak P_1^{e_1} \cdots \frak P_r^{e_r}.\]

The integer $e_i$ is called the
{\emph{ramification index}} of $\frak P_i$ over $\p$.  The {\emph{inertial
degree}} $f_i$ is defined to be $[\OO_K/\frak P_i : \OO_F/\p]$, the degree of the field extension $\OO_K/\frak P_i$ of $\OO_F/\p$.

When $K/F$ is a Galois extension, the ramification indices (resp. inertial degrees) are the same for all $i$ and we denote them simply by $e_\p$ (resp. $f_\p$). The following theorem describes the relationship between inertial and ramification degrees in this case.  

\begin{theorem}[{\cite[Theorem 21]{Mar}}] \label{pdec}
Let $K$ be a Galois extension of $F$ of degree $n$ and let $\p$ be a prime ideal of $\OO_F$. Then $\p \OO_K = ( \frak P_1 \cdots \frak P_{r_\p})^{e_\p}$ with $r_\p e_\p f_\p = n = [K:F]$.
\end{theorem}

In the special case of  cyclotomic extensions of $\Q$ we have the following result:
\begin{theorem}[{\cite[Theorem 26]{Mar}}] \label{markus26}
Let $F = \Q(\zeta_m)$, where $\zeta_m$ is a primitive $m$-th root of unity. Let $p$ be a prime not dividing $m$. Then the inertial degree $f_p$ in $F$ is the multiplicative order of $p$ modulo $m$. 
\end{theorem}

Finally, for the decomposition of primes in the extension $\Q(i)$ of $\Q$ (with $i^2=-1$) we obtain the following theorem as a special case of \cite[Theorem 25]{Mar}:

\begin{theorem}\label{pInZi}
Let $F =\Q(i)$ (and thus $\OO_F=\Z[i]$) and let $p$ be an odd prime. 

If $p \equiv 1 \mod 4$ then 
\[p\Z[i]= \p\bar\p  = (\pi)(\bar\pi),\]
where $\pi$ and $\bar\pi$ are non-associated, prime and of norm $p$ in $\Z[i]$. Thus $r_p=2$, $f_p=e_p=1$.

If $p \equiv 3 \mod 4$ then 
\[p\Z[i] = (p),\] 
i.e., the ideal $p\Z$ remains inert in $\Z[i]$ and $r_p=e_p=1$, $f_p=2$. 
\end{theorem}

\section*{Acknowledgment}
The authors wish to thank the referees for their constructive criticism of earlier versions of this paper.

\end{document}